\documentclass{PoS}

\usepackage{url}
\usepackage{array}
\usepackage{upgreek}

\newcommand{\dyb}{\textsc{Daya Bay}}
\newcommand{\ttt}{$\theta_{13}$}
\newcommand{\anue}{$\overline{\nu}_{e}$}
\newcommand{\dme}{$|\Delta \mathrm{m}^2_{ee}|$}

\title{Measurement of antineutrino oscillation with the full detector configuration at Daya Bay}

\ShortTitle{Measurement of antineutrino oscillation at Daya Bay}

\author{\speaker{Marco GRASSI}%
         \thanks{MG is supported by the Chinese Academy of Sciences President's International Fellowship Initiative, grant 2015PM007.}\\
        IHEP, Chinese Academy of Sciences, 19B Yuquanlu, Beijing, P.R. China\\
        E-mail: \email{mgrassi@ihep.ac.cn}\\
        on behalf of the \dyb{} Collaboration}

\abstract{In this poster, we present the latest measurement of electron antineutrino disappearance using the fully constructed
 Daya Bay Reactor Neutrino Experiment. 
 A total exposure of $6.9 \times 10^5$ GW$_{\mathrm{th}}$ ton days was achieved in November 2013 after
 617 day of data taking.
 The most precise estimates to date of the neutrino mass and mixing parameters 
 $|\Delta \mathrm{m}^2_{ee}|$ and $\sin^2 2 \theta_{13}$  was obtained with an analysis
of the relative antineutrino rates and energy spectra between detectors.
 The value of the two parameters was found to be  $\sin^2 2 \theta_{13} = 0.084 \pm 0.005$ and
 $|\Delta \mathrm{m}^2_{ee}| = (2.42 \pm 0.11) \times 10^{-3}\,\mathrm{eV}^2$.
  This poster focuses in particular on describing how improvements in the calibration and in the 
 energy response model contributed to achieve this result.}
\FullConference{XXVII International Symposium on Lepton Photon Interactions at High Energies\\
		17-22 August 2015\\
		Ljubljana, Slovenia}

\begin{document}

\section{Experimental Apparatus and Methods}

Neutrino flavour oscillation due to the mixing angle \ttt{} has been observed using reactor 
antineutrinos (\anue{}) \cite{cit1, cit2, cit3} and accelerator neutrinos~\cite{cit4,cit5}. The 
\dyb{} experiment previously reported the discovery of a non-zero value of $\sin^2 2\theta_{13}$ by observing the disappearance of reactor 
antineutrinos over kilometre distances~\cite{cit1}, and the first measurement of the effective mass splitting 
$|\Delta m^{2}_{ee}|$ via the distortion of the \anue{} energy spectrum~\cite{cit9}. 
This poster presents new results with larger statistics and 
significant improvements in energy calibration and background reduction published in \cite{cit0}.

The \dyb{} experiment consists of eight functionally identical antineutrino detectors (ADs) hosted in three underground experimental
halls (EHs), detecting reactor \anue{} via inverse beta decay (IBD) reactions. 
EH1 and EH2 are  respectively located at short distance from the Daya Bay and Ling Ao reactor cores, 
and they both host two ADs. EH3 is at 1.6~km distance from the cores and hosts 4 ADs. Each EH is equipped with a 
muon detector system to tag and veto cosmic muons, consisting of a layer of resistive plate chambers (RPCs)
and a water Cherenkov detector, in which the ADs are immersed.
%and an instrumented water pool (in which the ADs are immersed) to detect Cherenkov light.

Each AD has three nested cylindrical volumes separated by concentric acrylic vessels, while the outermost vessel is made of stainless steel.
The innermost volume holds 20~ton of gadolinium-doped liquid scintillator (Gd-LS) that serves as the antineutrino target. 
The middle volume is called the gamma catcher and is filled with 21~ton of undoped liquid scintillator (LS) for detecting gamma rays that escape 
the target volume. 
%The gamma catcher increases the containment of gamma energy, thus improving the energy resolution and reducing the 
%uncertainties of the antineutrino detection efficiency. 
The outer volume contains 37~ton of mineral oil (MO) to provide optical homogeneity 
and to shield the inner volumes from background radiation. 
The outer volume also hosts 192 eight-inch PMTs facing the target volume.

IBDs are selected by exploiting the prompt-delayed time structure of the event, where the positron annihilates soon after its production 
(prompt energy), while the neutron gets captured on gadolinium with a mean capture time of $\sim$30~$\upmu \mathrm{s}$ (delayed energy). 
In order for an event to be selected, the difference between the two energy depositions is required to be between 1 and 200~$\upmu \mathrm{s}$.

\noindent\begin{tabular*}{\textwidth}{ p{0.43\textwidth} p{0.02\textwidth} p{0.45\textwidth}   }
\includegraphics[height=4.9cm]{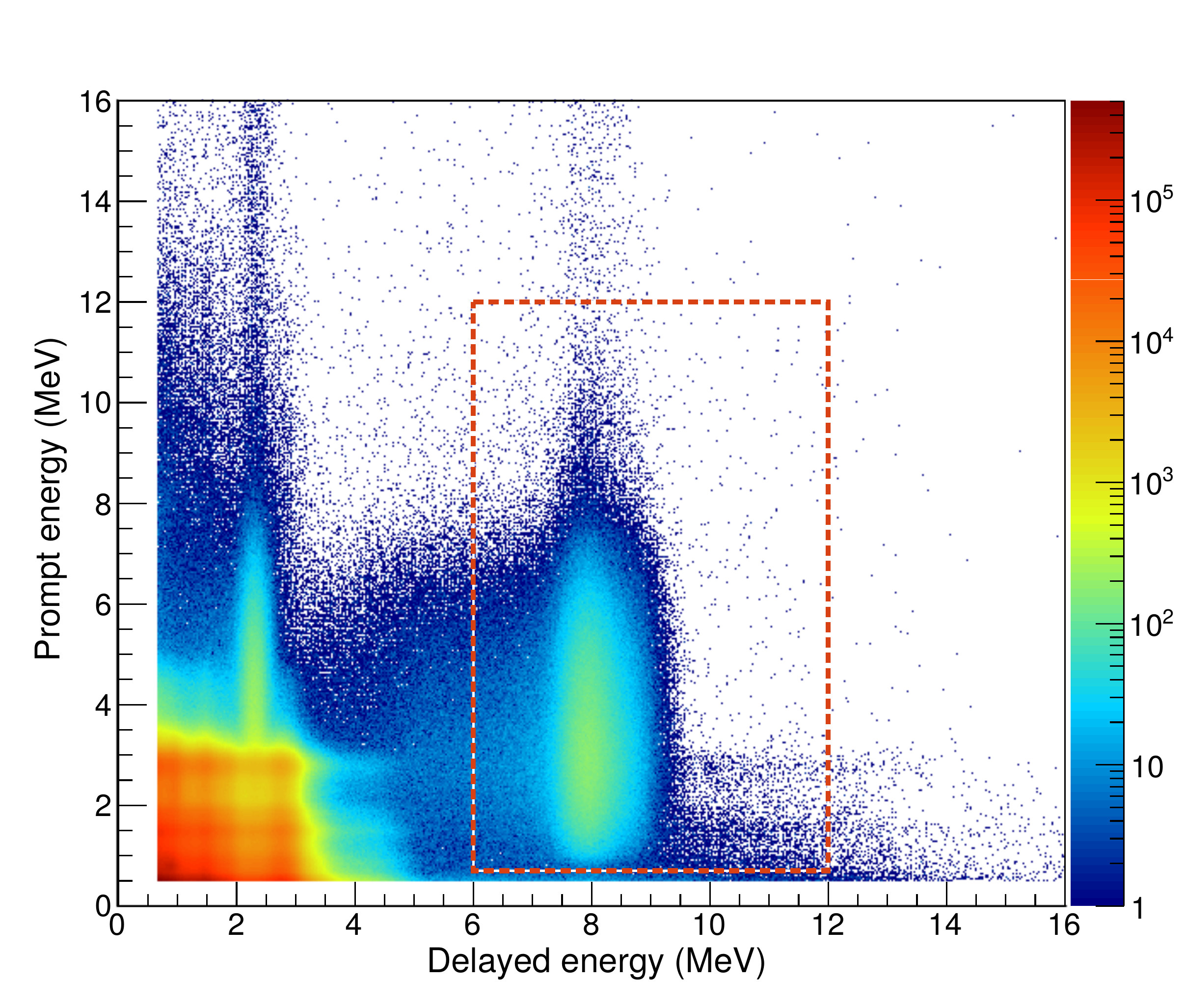}
& &
\includegraphics[height=4.9cm]{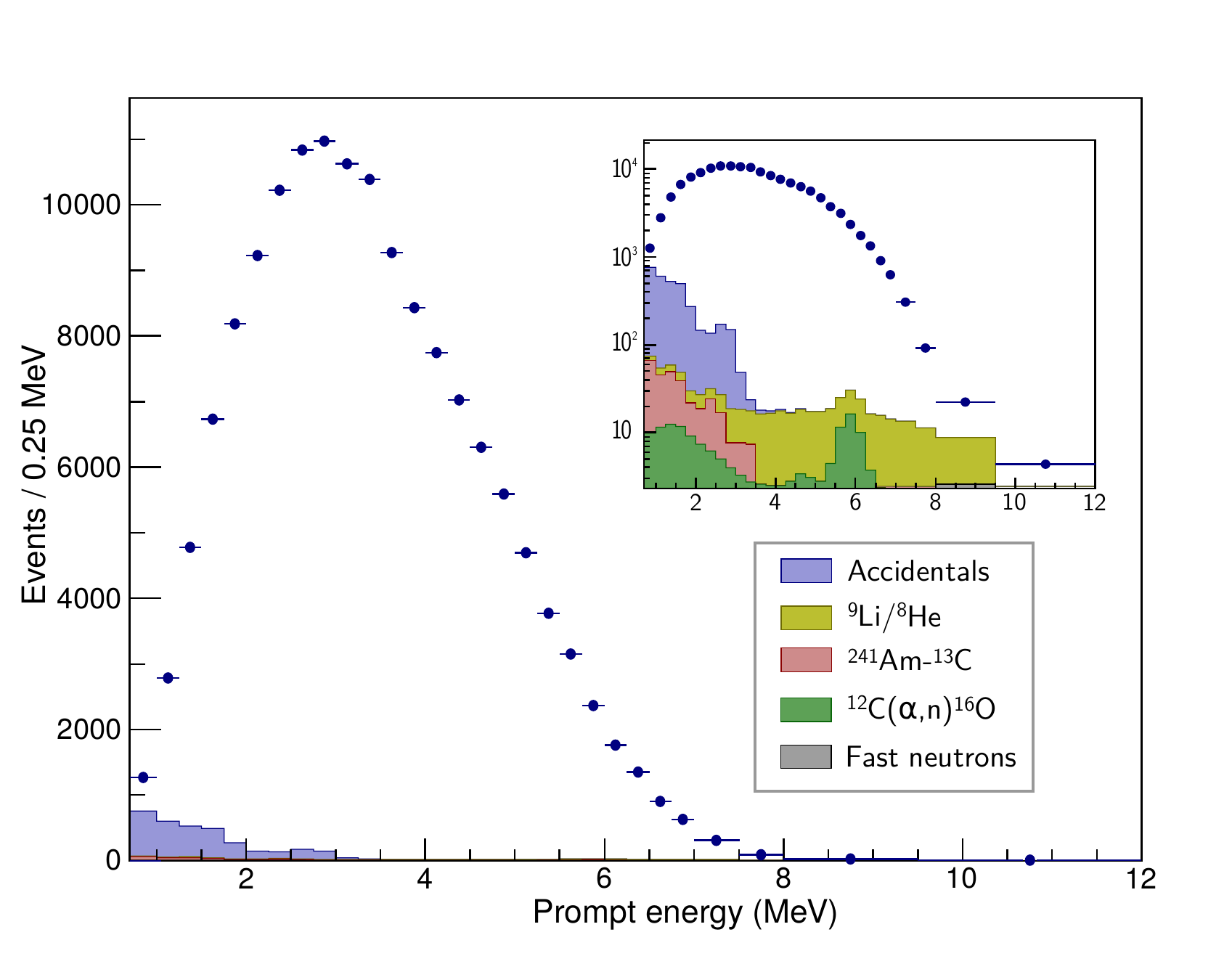} \\
{\small \textbf{Figure 1.} IBD prompt/delayed energy spectrum. The dashed line shows the energy-based selection criteria.}
& &
{ \small \textbf{Figure 2.} Prompt energy spectrum of IBD candidates in EH3 together with the subtracted background components. 
The inset shows the same plot in  logarithmic scale.\vspace{10pt}}\\
\end{tabular*}

\noindent\begin{tabular*}{\textwidth}{ p{0.45\textwidth} b{0.45\textwidth}   }
 \includegraphics[height=5.8cm]{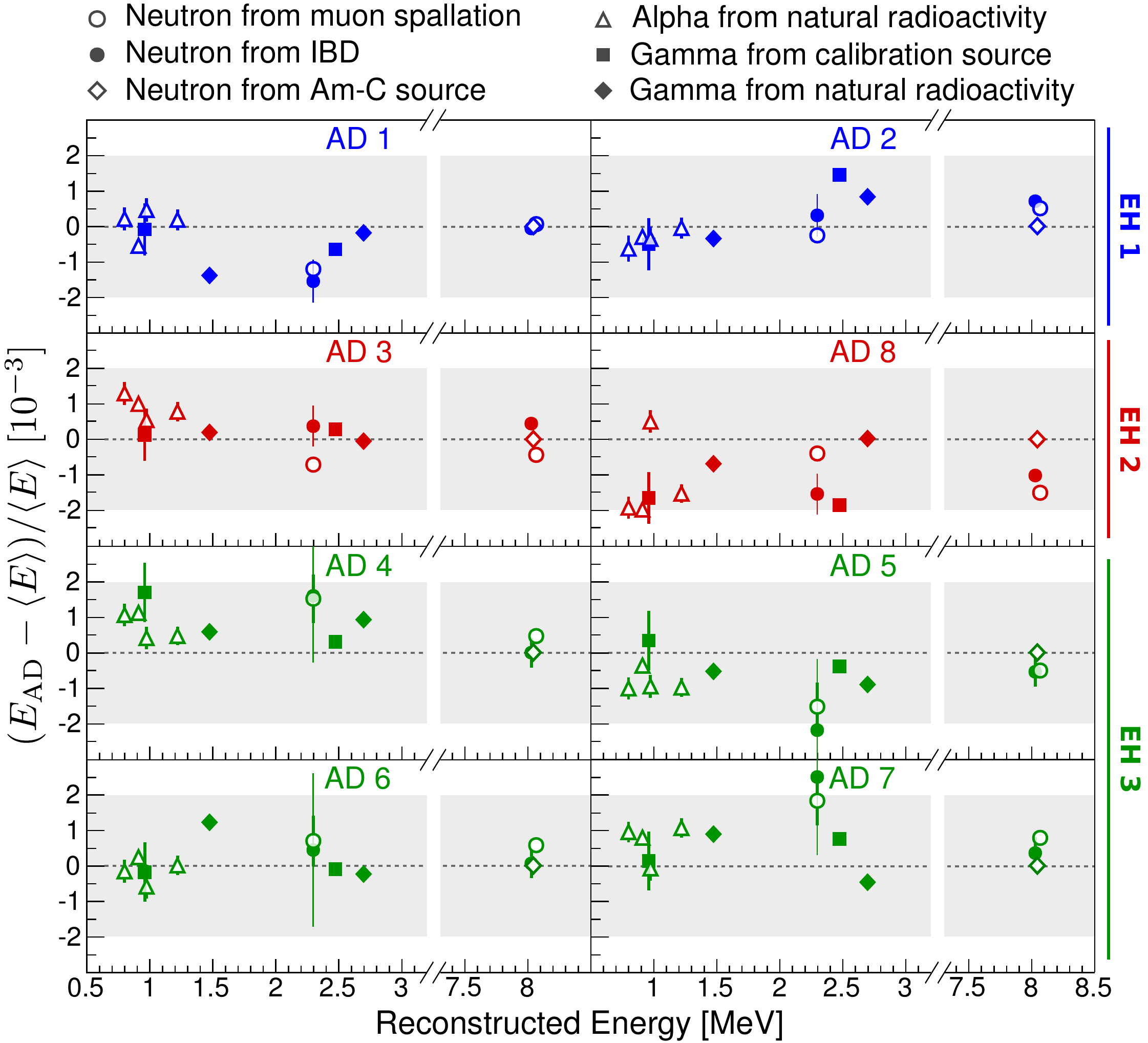} \vspace{10pt}
& 
{\vspace{15pt} \small \textbf{Figure 3.} Comparison of the reconstructed energy between antineutrino detectors for a variety of calibration references. 
$\mathrm{E_{AD}}$ is the reconstructed energy determined using each AD, and $\langle E \rangle$ is the 8-detector average. 
Error bars are statistical only, and systematic variations between detectors for all calibration references are \textless0.2\%. 
The \textasciitilde8 MeV n-Gd capture gamma peaks from Am-C sources are used to define the energy scale of 
each detector, and hence show zero deviation.} \\
\end{tabular*}

\noindent Additional energy selection criteria require the prompt energy to be in the 0.7-12~MeV range, and the delayed energy
to be in the 6-12~MeV range, as shown in Fig.~1.
In order to suppress cosmogenic products, IBD candidates are rejected if the delayed signal occurs
soon after a muon trigger, where the vetoed time window can range from 600~$\upmu$s to 1~s according the 
amount of energy deposited in the detector.
 The prompt energy spectrum of selected events, together with the five major sources of background
 (fast neutrons of cosmogenic origin, correlated $\beta$-n decays from cosmogenic $^9$Li and $^8$He,
 $^{13}\mathrm{C}(\alpha , \, \mathrm{n}) ^{16}\mathrm{O}$ reactions, single neutrons from the Am-C  calibration sources
 and accidental coincidences), are shown in Fig.~2.

We use the prompt energy as a proxy for the incident \anue{} energy, and any difference in the energy response between ADs
affects the estimation of \dme{}. 
The detector energy scale is calibrated using Am-C neutron sources deployed at the detector centre, 
with the \textasciitilde8~MeV peaks from neutrons captured on Gd aligned across all eight ADs. 
The time variation and the position dependence of the energy scale are corrected using the 2.5~MeV gamma-ray peak from 
$^{60}$Co calibration sources. 
The uncertainty associated to any residual difference in the energy response is evaluated by comparing several
calibration reference points across all ADs, as shown in Fig.~3. 
%
%The spatial distribution of each calibration reference varies, incorporating deviations in spatial response between detectors.
%
Such reference points are:
%
%The reconstructed energies of various calibration reference points in different ADs are compared in Fig.~3. 
%The residual energy scale 
%Fig.~3 presents measurements of 
(\textsc{i}) $^{68}$Ge, $^{60}$Co and Am-C calibration sources placed at detector centre,
(\textsc{ii}) Gd-captured neutrons from IBD and muon spallation, distributed nearly uniformly throughout the Gd-LS region, 
%References distributed inside and outside of the target volume are 
(\textsc{iii}) neutrons being captured on $^1$H, intrinsic $\alpha$ particles from polonium and radon decays, and gammas from $^{40}$K and $^{208}$Tl decays, all being distributed inside and outside of the target volume. 
Events from all the calibration samples are required to have the reconstructed vertex within the Gd-LS region. 
%Events from all the calibration samples are selected 
However, since  the spatial distribution of each calibration sample varies, the energy difference shown in Fig.~3
incorporates also deviations in spatial response between detectors.   The resulting uncorrelated relative uncertainty of the energy scale 
is 0.2\%.

The detector energy response is known to be non-linear. The main causes of this behaviour are: (\textsc{i}) a particle-dependent 
non-linear light yield of the scintillator, and (\textsc{ii}) a charge-dependent non-linearity in the PMT readout electronics, 
each being at the level of 10\% within the IBD energy range. 
We model the non-linear response with a semi-empirical model, which can be factorised into two independent components: 
$f_{\mathrm{scintillator}}(k_B, k_C)$ ---where the two free parameters are the Birks' constant $k_B$ and
the fraction of Cherenkov light contributing to the total light yield $k_C$---, and $f_{\mathrm{electronics}}(\alpha, \tau)$ ---where 
$\alpha$ and $\tau$ are respectively the amplitude and the scale of the exponential describing the electronics non-linearity.

\noindent\begin{tabular*}{\textwidth}{ p{0.45\textwidth} p{0.02\textwidth} p{0.45\textwidth}   }

\includegraphics[height=5cm]{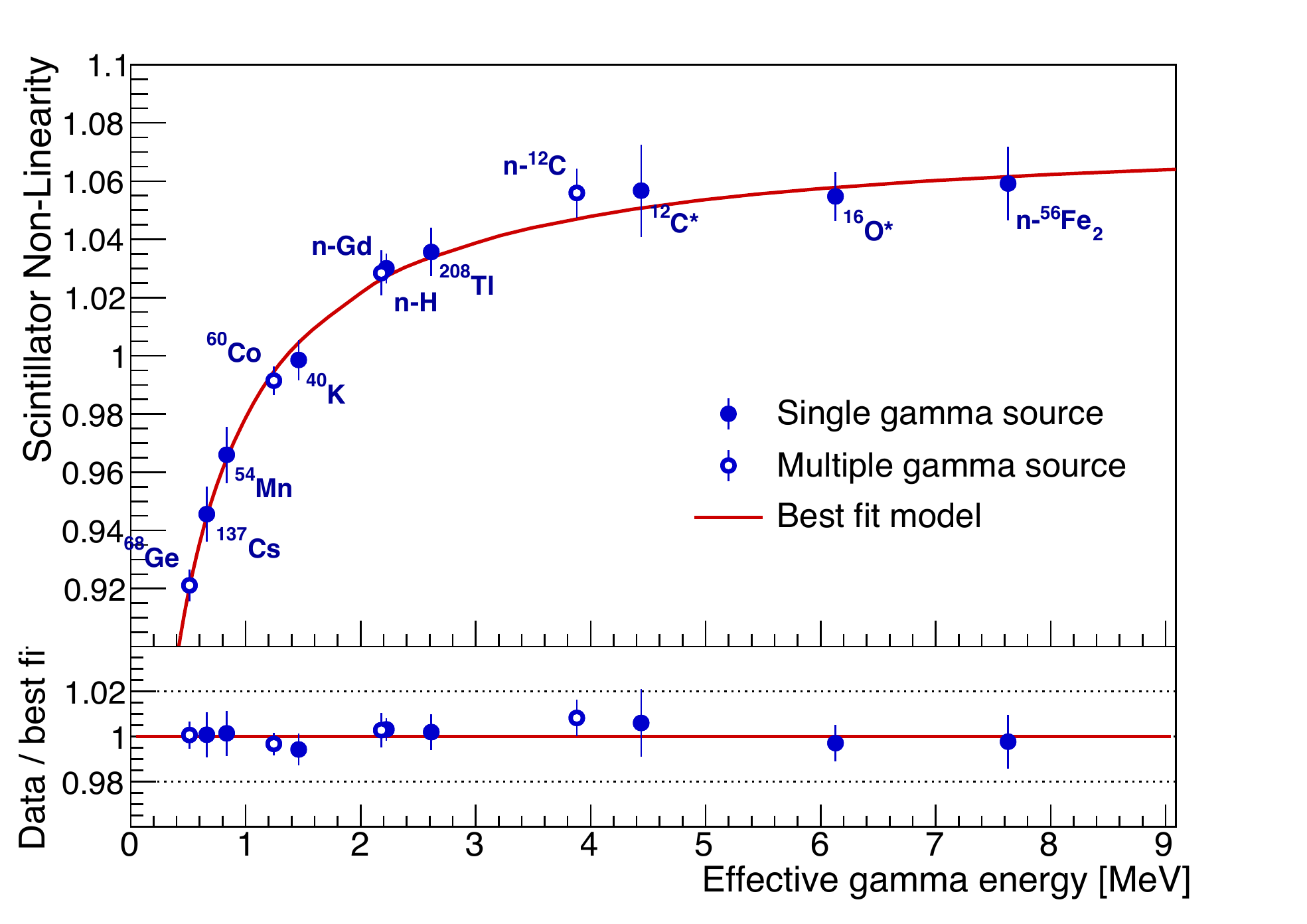}
& &
\includegraphics[height=5cm]{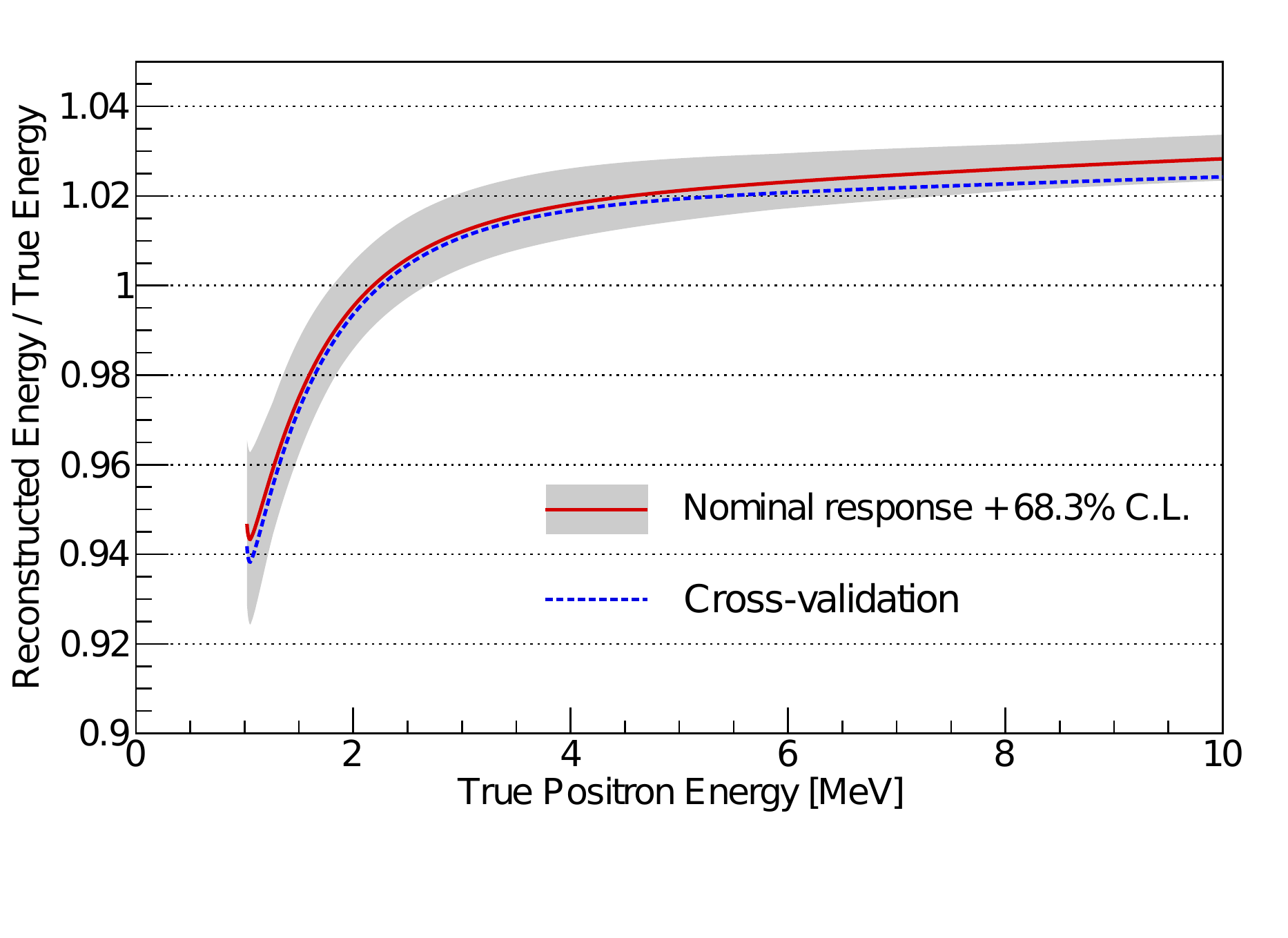} \\
{\small \textbf{Figure 4.} The scintillator non-linearity resulting from the best-fit parameters of the overall energy response model,
compared to deployed and intrinsic  gamma calibration sources.}
& &
{ \small \textbf{Figure 5.} \textit{Solid curve}: Estimated energy response of the detectors to positrons,
based on gamma rays from both deployed and intrinsic sources, as well as electrons from $^{12}$B $\beta$ decays.
\textit{Dashed curve}: Validation model based on 
$^{12}$Bi, $^{214}$Bi, $^{208}$Tl $\beta$\texttt{+}$\gamma$ spectra together with the
53-MeV edge in the Michel electron spectrum. \vspace{10pt}}\\
\end{tabular*}

\noindent The values of these four free parameters are obtained from an unconstrained $\chi^2$ fit to several calibration datasets, 
namely 12 gamma lines by both artificial sources deployed at detector centre and naturally occurring sources, and
the continuous $\beta$ decay spectrum of $^{12}$B resulting from muon spallation events in the Gd-LS volume. 
The comparison of the best fit model against gamma data is shown in Fig.~4, while
the nominal positron response derived from the best fit parameters is represented by the solid curve in Fig.~5. 
The depicted uncertainty band represents other response functions
consistent with the fitted calibration data within a 68.3\% C.L.
The positron response is further validated 
with
%by fitting the same model to 
the 53-MeV cutoff in the decay electron 
spectrum from muon decay at rest, and with the continuous $\beta$\texttt{+}$\gamma$ spectra from natural bismuth and thallium decays. 
This additional model ---represented by the blue dashed curve in Fig.~5--- falls within the 1-$\sigma$ contour of the nominal 
positron response curve, improving our confidence in the characterization of the absolute energy response of the detectors.

\section{Results}

Neutrino oscillation parameters are measured using the L/E-dependent disappearance of \anue{}, as given by the survival probability
\begin{equation}
P \simeq 1- \cos^4 \theta_{13} \, \sin^2 2 \theta_{12} \, \sin^2 \frac{1.267 \Delta m_{21}^2 \, L }{E} -  \sin^2 2 \theta_{13} \, \sin^2 \frac{1.267 \Delta m^2_{ee} \, L }{E} \, \, .
\end{equation}
Here E is the energy in MeV of the \anue{}, L is the distance in meters from its production point, 
$\theta_{12}$ is the solar mixing angle, and $\Delta m^2_{12} = m_2^2 - m_1^2$ is the mass-squared 
difference of the first two neutrino mass eigenstates in eV$^2$.
Since recent measurements of the IBD positron energy spectrum disagree with models of reactor \anue{} 
emission~\cite{cit3, cit20, cit21}, 
to measure the neutrino oscillation parameters we employ a technique predicting the signal in the far hall based on 
measurements obtained in the near halls.
This allows us to minimise the result's dependence on models of the reactor antineutrino flux
(more information can be found in \cite{cit0}). Out of this approach, we obtain $\sin^2 2 \theta_{13} = 0.084 \pm 0.005$
and $\Delta \mathrm{m}^2_{ee} = (2.42 \pm 0.11) \times 10^{-3}\,\mathrm{eV}^2$, with $\chi^2$/NDF = 134.6/146.
%To compute such values we use $\sin^2 2 \theta_{12} = 0.857 \pm 0.024$ and 
%$\Delta m^2_21 = (7.50 \pm 0.20) \times 10^{-5} \, \mathrm{eV}^2$ from \cite{cit31}, but our result is largely 
%independent of this external input. 
Under the normal (inverted) hierarchy assumption, \dme{} yields
$\Delta m_{32} = (2.37 \pm 0.11) \times 10^{-3} \mathrm{eV}^2$
($\Delta m_{32} = -(2.47 \pm 0.11) \times 10^{-3} \mathrm{eV}^2$).
This result is consistent with and of compatible precision to measurements obtained  from accelerator
$\nu_{\mu}$ and $\overline{\nu}_{\mu}$ disappearance~\cite{cit10, cit11}.
The reconstructed positron energy spectrum observed in the far site is compared in Fig.~6 with the 
expectation based on the near-site measurements.
%The spectral shape from all experimental halls is compared in Fig.~7 to the electron antineutrino survival probability 
%assuming our best estimates of the oscillation parameters. 
The total uncertainties of both 
$\sin^2 2 \theta_{13}$ and \dme{} are dominated by statistics. The most significant systematic uncertainties for 
$\sin^2 2 \theta_{13}$ are due to the relative detector efficiency, reactor power,
relative energy scale, and $^9$Li/$^8$He background.
The systematic uncertainty in \dme{} is dominated by uncertainty in the relative energy scale.

\noindent\begin{tabular*}{\textwidth}{ p{0.45\textwidth} p{0.02\textwidth} b{0.45\textwidth}   }
\includegraphics[height=7cm]{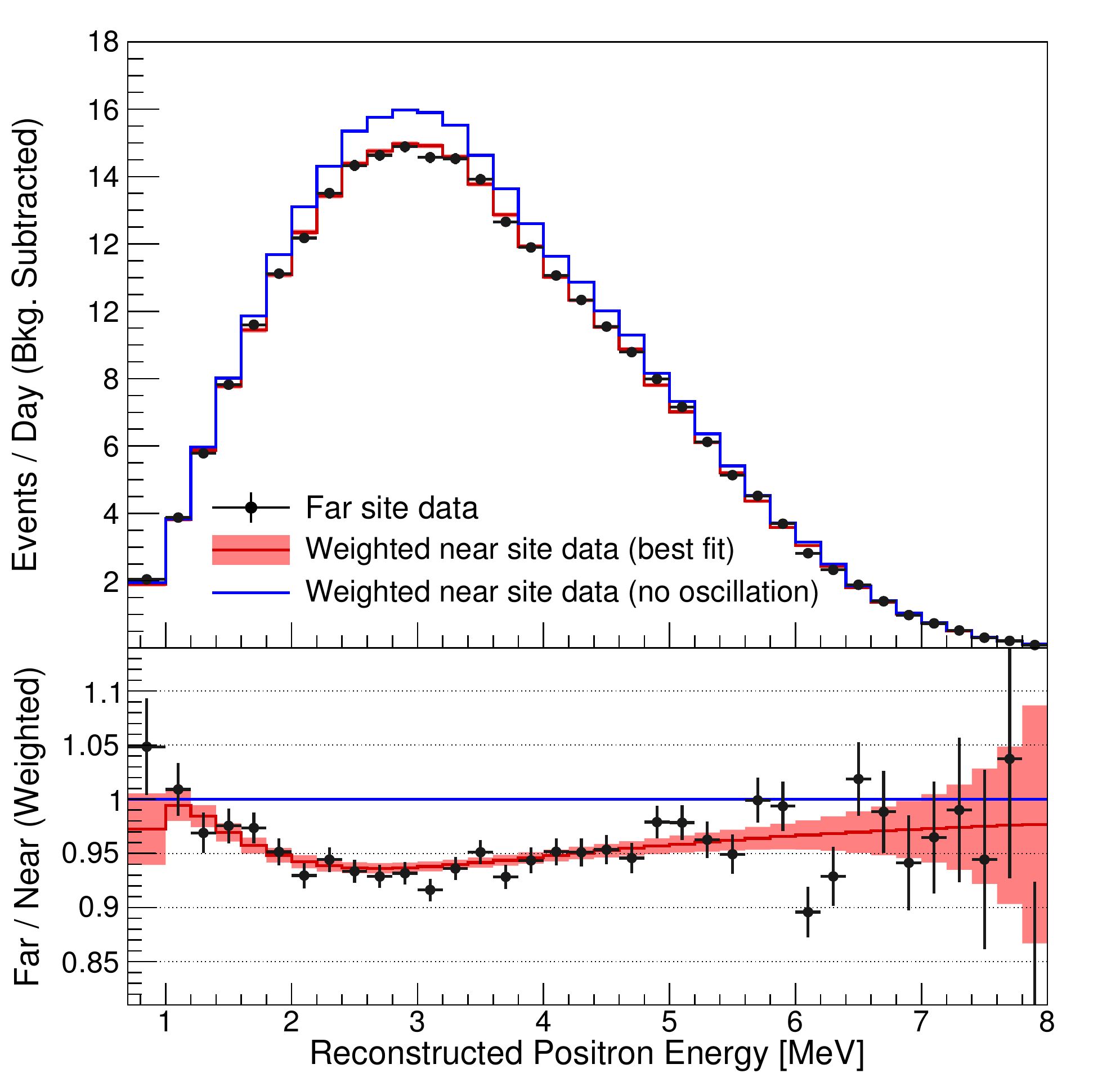}
& &
{\small \textbf{Figure 6.} \textit{Top}: Background-subtracted reconstructed positron energy spectrum observed in the far site (black points), 
as well as the expectation derived from the near sites excluding (blue line) or including (red line) our best estimate of oscillation. 
The spectra were efficiency corrected and normalised to one day of live time. 
\textit{Bottom}: Ratio of the spectra to the no-oscillation case. 
The error bars show the statistical uncertainty of the far site data. 
The shaded area includes the systematic and statistical uncertainties from the near-site measurements.}\\ 
%
%hE?i was calculated for each bin using the estimated detector response, and Leff was obtained by equating the actual flux to an effective antineutrino flux using a single baseline.}\\
\end{tabular*}

%\noindent\begin{tabular*}{\textwidth}{ p{0.45\textwidth} p{0.02\textwidth} p{0.45\textwidth}   }
%\includegraphics[height=5cm]{img/spectral_distortion}
%& &
%\includegraphics[height=5cm]{img/loe} \\
%{\small \textbf{Figure 6.} \textit{Top}: Background-subtracted reconstructed positron energy spectrum observed in the far site (black points), 
%as well as the expectation derived from the near sites excluding (blue line) or including (red line) our best estimate of oscillation. 
%The spectra were efficiency corrected and normalised to one day of live time. 
%\textit{Bottom}: Ratio of the spectra to the no-oscillation case. 
%The error bars show the statistical uncertainty of the far site data. 
%The shaded area includes the systematic and statistical uncertainties from the near-site measurements.}
%& &
%{ \small \textbf{Figure 7.} \anue survival probability 
%versus effective propagation distance $L_{\mathrm{eff}}$ divided by the average antineutrino energy $\langle E_{\nu}\rangle$. 
%The data points represent the ratios of the observed antineutrino spectra to the expectation assuming no oscillation. 
%The solid line represents the expectation using the best estimates of $\sin^2 2 \theta_{13}$ \dme{}. The error bars are statistical only.}\\ 
%\end{tabular*}


\begin{thebibliography}{9}



\bibitem{cit1}
F.~P.~An {\it et al.} [Daya Bay Collaboration],
Phys.\ Rev.\ Lett.\  {\bf 108} (2012) 171803.

\bibitem{cit2}
  J.~K.~Ahn {\it et al.} [RENO Collaboration],
  Phys.\ Rev.\ Lett.\  {\bf 108} (2012) 191802.

\bibitem{cit3}
  Y.~Abe {\it et al.} [Double Chooz Collaboration],
  %``Improved measurements of the neutrino mixing angle $\theta_{13}$ with the Double Chooz detector,''
   JHEP 1410 (2014) 086 [err: JHEP {\bf 1502} (2015) 074].

\bibitem{cit4}
  K.~Abe {\it et al.} [T2K Collaboration],
  %``Observation of Electron Neutrino Appearance in a Muon Neutrino Beam,''
  Phys.\ Rev.\ Lett.\  {\bf 112} (2014) 061802

\bibitem{cit5}
  P.~Adamson {\it et al.} [MINOS Collaboration],
  %``Electron neutrino and antineutrino appearance in the full MINOS data sample,''
  Phys.\ Rev.\ Lett.\  {\bf 110} (2013) 17,  171801

%\bibitem{cit6}
%F.~P.~An {\it et al.} [Daya Bay Collaboration],
%Chin.\ Phys.\ C {\bf 37} (2013) 011001.
  
%\bibitem{cit7}
%F.~P.~An {\it et al.} [Daya Bay Collaboration],
%Phys.\ Rev.\ D {\bf 90} (2014) 7,  071101.

\bibitem{cit9}
F.~P.~An {\it et al.} [Daya Bay Collaboration],
Phys.\ Rev.\ Lett.\  {\bf 112} (2014) 061801.

  \bibitem{cit0}
F.~P.~An {\it et al.} [Daya Bay Collaboration],
Phys.\ Rev.\ Lett.\  {\bf 115} (2015),  111802.

\bibitem{cit20}
F.~P.~An {\it et al.} [Daya Bay Collaboration],
arXiv:1508.04233.

\bibitem{cit21}
S. Seo (RENO Collaboration), in Neutrino2014: 
The XXVI International Conference on Neutrino Physics and Astrophysics, Boston, 2014 (unpublished).

\bibitem{cit31}
J. Beringer et al. (Particle Data Group), Phys. Rev. D 86, 010001 (2012), see Sec. 13.

\bibitem{cit10}
P. Adamson et al. (MINOS Collaboration), Phys. Rev. Lett. 112, 191801 (2014).

\bibitem{cit11}
K. Abe et al. (T2K Collaboration), Phys. Rev. Lett. 112, 181801 (2014).

\end{thebibliography}
\end{document}